\documentclass[aps,12pt,superscriptaddress,amsfonts,amssymb,amsmath]{revtex4}
\pdfoutput=1

\usepackage{graphicx}
\usepackage{epsfig}
\usepackage{makeidx}
\usepackage{epstopdf}
\usepackage{natbib}
\usepackage{xcolor}

\begin{document}

\title{Quantum-only metrics in spherically symmetric gravity}

\author{G.\ Modanese \footnote{Email address: giovanni.modanese@unibz.it}}
\affiliation{Free University of Bozen-Bolzano \\ Faculty of Science and Technology \\ I-39100 Bolzano, Italy}

\linespread{0.9}

\begin{abstract}

The Einstein action for the gravitational field has some properties which make of it, after quantization, a rare prototype of systems with quantum configurations that do not have a classical analogue. Assuming spherical symmetry in order to reduce the effective dimensionality, we have performed a Monte Carlo simulation of the path integral with transition probability $e^{-\beta |S|}$. Although this choice does not allow to reproduce the full dynamics, it does lead us to find a large ensemble of metric configurations having action $|S|\ll \hbar$ by several magnitude orders. These vacuum fluctuations are strong deformations of the flat space metric (for which $S=0$ exactly). They exhibit a periodic polarization in the scalar curvature $R$. In the simulation we fix a length scale $L$ and divide it into $N$ sub-intervals. The continuum limit is investigated by increasing $N$ up to $\sim 10^6$; the average squared action $\langle S^2 \rangle$ is found to scale as $1/N^2$ and thermalization of the algorithm occurs at a very low temperature (classical limit). This is in qualitative agreement with analytical results previously obtained for theories with stabilized conformal factor in the asymptotic safety scenario. 

\end{abstract}

\maketitle

\section{Introduction}

Efforts towards the unification of General Relativity and Quantum Mechanics into a coherent theory of Quantum Gravity were started long ago and have been intensifying in the last decades. In spite of big progress in loop Quantum Gravity \cite{rovelli2004quantum,rovelli2014covariant}, asymptotic safety \cite{reuter2018quantum} and discrete spacetime models \cite{hamber2008quantum,hamber2019vacuum,ambjorn2012nonperturbative,loll2019quantum}, ``the revolution is still unfinished'', in the words of C.\ Rovelli. The spin-offs of this research work, however, are manifold and remarkable in their own right. In general, it is fair to say that the quest for unification has led to a better comprehension of both General Relativity and Quantum Mechanics. 

We made some early contributions to Quantum Gravity in the covariant formulation by showing that the Wilson loop vanishes to leading order \cite{modanese1994wilson} and proposing an alternative expression for the static potential of two sources \cite{modanese1995potential}. This formula was used by Muzinich and Vokos \cite{muzinich1995long} and by Hamber and Liu \cite{hamber1995quantum}, respectively in perturbation theory and in non-perturbative Regge calculus, to give an estimate of quantum corrections to the Newton potential. The vanishing of the Wilson loop (and of curvature correlations \cite{modanese1992vacuum}) to leading order is only one of the peculiar aspects of the quantum field theory of gravity, that sets it apart from other successful quantum field theories like QED and QCD.

One of the problems with defining quantum theories of gravity is that, unlike for other quantum mechanical systems, there is no action or Hamiltonian that would be bounded from below. As a result any formal definition of a path integral or partition function would be plagued by divergences and be dominated by configurations which have arbitrarily negative energy. This paper reports on a study in this context, where we only restrict to spherically symmetric configurations. The goal is to identify configurations with zero or almost zero action, which would all equally contribute to a quantum path integral (being unsuppressed by an action factor). This is done numerically and the paper reports on results as the number of steps in the discretisation is increased.

Although many possible extensions and generalizations of the Einstein action have been proposed \cite{nojiri2017modified}, which could help in addressing open issues in cosmology, the Einstein action is the natural action at intermediate energies and arises directly from the quantization of massless spin 2 fields.

In our opinion, the indefinite sign of the Einstein action gives us a chance to explore a phenomenon that is otherwise unknown in quantum field theory and more generally in Quantum Mechanics, namely the existence of configurations for which the action is zero, like for the classical vacuum, but not a stationary point. We call them zero modes of the action and we have proven analytically their existence for the Einstein action as well as for a peculiar elementary quantum system (the massless harmonic oscillator \cite{modanese2016functional,modanese2017ultra}).

The purpose of this work is to show that if the condition $S=0$ defining the zero modes is relaxed to $S/\hbar \ll 1$, then a large ensemble of these modes can be numerically constructed via a Metropolis -- Monte Carlo algorithm. The condition $S/\hbar \ll 1$ implies that these modes can play an important role in the path integral, although they are very different from classical solutions. In this sense the Einstein action offers an example of a dynamical system with unique quantum properties and a possible prototype for similar systems in other branches of physics.

The algorithm for the generation of the zero modes ensemble has been presented in \cite{modanese2019metrics}, but the application was limited to metric configurations at the Planck scale, and accordingly the discretization limited to $N=10^2$ space sub-intervals. 
Several authors have found, with various techniques, that the vacuum state of quantum gravity has a non-trivial structure at that scale. In particular, field configurations with spherical symmetry have been considered by \cite{preparata2000gas,garattini2002spacetime}. One may wonder how this structure scales up to larger distances, and that is the main purpose of this work. After returning to more transparent physical units, we have performed several simulations looking for quantum zero modes with $S/\hbar \ll 1$ at a scale $L \gg L_P$ and in the continuum limit $N\to \infty$. It turns out that just the continuum limit allows to obtain such modes. The exact scaling dependence on $L$ and $N$ is reported in Sect.\ \ref{sec-scaling}.

Our results, though obtained in a different setting and with different methods, appear to be close to what is found in a paper by Bonanno and Reuter (\cite{bonanno2013modulated}; see also \cite{bonanno2019structure}).
In this work, the authors have added an $R^2$ term to the action to make it bounded from below, and they find indications for a ground state which violates translational symmetry and displays a ``rippled'' structure.
They argue that a ``kinetic condensate'' characterizes the vacuum state of asymptotically safe quadratic gravity theories, so that if this scenario is realized in the full theory, the vacuum state of gravity is the gravitational analogous to the Savvidy vacuum in Quantum Chromo-Dynamics.
A more detailed comparison between our results and those of \cite{bonanno2013modulated} will be given in Sect.\ \ref{sec-limits}. 

The outline of the paper is the following. In Sect.\ \ref{sec-dis} we recall the form of the Einstein action reduced for spherically symmetric and time-independent metrics, first in the continuum version and then in the discretized version. We also recall our previous results from simulations at the Planck scale, made using a small number $N$ of sub-intervals. In Sect.\ \ref{sec-scaling} we analyze the scaling properties of the discretized action with respect to $N$, up to $N\sim 10^6$, and also the scaling with respect to the inverse temperature $\beta$ of the Metropolis algorithm and the length scale $L$ of the vacuum fluctuations. In Sect.\ \ref{pol} the observed polarization patterns of the metrics are reported and discussed. Sect.\ \ref{sec-2di} offers for illustration purposes a mathematical example of zero modes in an oscillating 2D integral. Sect.\ \ref{sec-ext} considers a possible extension to higher dimensions. Finally, Sect.\ \ref{sec-concl} briefly summarizes our conclusions.

\section{The discretized action}
\label{sec-dis}

Our physical model \cite{modanese2019metrics} is defined by the Einstein action of the gravitational field, computed for a metric with spherical symmetry and independent from time. The only field variable is the metric component $g_{rr}(r) = A(r)$, with $0 \le r < +\infty$. This kind of dimensional reduction of gravity has been already employed in several classical and quantum models.

The continuum action is
\begin{equation}
S=\frac{\tau}{16\pi G} \int_0^\infty dr \sqrt{A}\left( \frac{rA'}{A^2}+1-\frac{1}{A}\right)
\label{S-cont}
\end{equation}

This is derived from the usual expression of the Einstein action $\int \sqrt{g}Rd^4x$ in units such that $c=1$, and in which the integral over time has been replaced by a factor $\tau$, meaning that the metrics we are considering are stationary but have a limited duration $\tau$. This is clearly an approximation, and eventually we should introduce a function of time describing an adiabatic switch on/off; we expect the corresponding time derivatives to give a negligible contributions to the curvature for the values of $\tau$ and $L$ considered in this paper. (This can be checked for simplicity in the linearized approximation, where the scalar curvature is simply given by $R_{lin}=\partial_\mu \partial^\mu h_\nu^\nu-\partial_\mu \partial_\nu h^{\mu \nu}$; with the chosen form of the metric, the only time-dependent contribution is $\partial_0^2 h_{rr} \propto \tau^{-2} h_{rr} \ll \partial_r^2 h_{rr}$.)

In principle it is possible to improve the model by increasing the number of degrees of freedom, making the algorithms more complicated but still manageable, at least in two ways: (1) besides the component $g_{rr}$, consider as variable also the component $g_{00}(r)$, which at the moment is taken constant and equal to 1; (2) introduce a dependence on an angle $\theta$. The full expression of $R$ in this case is given in \cite{modanese2019metrics} and refs.

In the discretized version of the action the variable $r$ runs on an interval $(0,L)$ divided into $N$ parts. After defining $\delta=L/N$, we can say that $r$ takes the values 0, $\delta$, $2\delta$, ... , $N\delta$, or $\{h\delta, h=0,1,...,N\}$.
The field takes values $A_h$, $h=0,1,...,N$, corresponding to $A(h\delta)$.

The boundary condition on the right end of the interval is $A(L)=A_N=1$, while on the left, for $A(0)$, we do not set any constraint.
We suppose that $A(r)=1$ for $r \ge L$. As a consequence, we are not considering metric perturbations extended to infinity, but only fields different from flat space in $(0,L)$ (localized fluctuations). Clearly, $L$ must be regarded as one of the parameters for which we will need a scaling analysis.

Upon quantization the variables $A_0$, $A_1$, ... $A_N$ become the integration variables of a path integral, with measure given by the DeWitt super-metric (see \cite{hamber2008quantum}, Sect.\ 2.4). Since the action is not positive definite, we suppose at the beginning that this path integral is of the Lorentzian kind, with weight $e^{iS/\hbar}$.

In the continuum action (\ref{S-cont}) we replace the integral with a sum and so we obtain the discretized action
\begin{equation}
S\simeq \frac{\tau L}{GN} \sum_{h=0}^N S_h
\label{S-disc}
\end{equation}
with 
\[
S_h=\sqrt{|\hat{A}_h|} \left( \frac{2h+1}{2\hat{A}_h^2} (A_{h+1}-A_h)+1-\frac{1}{\hat{A}_h} \right)
\]
and
\[
\hat{A}_h=(A_{h+1}+A_h)/2
\]

We are looking for anomalous fluctuations with respect to the trivial classical solution $A(r)=1$ everywhere, which gives $S=0$ (flat space). Our idea is to use the path integral as follows: if there is an ensemble of non-trivial metrics such that $S/\hbar \ll 1$, we suppose that they may describe important vacuum fluctuations. They do not need to be stationary points of the action like the classical configurations; they can also be exact zero modes of the action (for example, those we have found already with analytical techniques [CQG]) or modes with almost-zero action. An important requirement to make them relevant is that they must have a large volume in configuration space: the Montecarlo simulations will tell us if this is the case, and we may also expect (as confirmed in \cite{modanese2019metrics}) that according to the same simulations certain exact analytical zero modes will turn out to be too little probable to be physically relevant. A further discussion of this idea in relation to the stationary phase principle can be found in Sect.\ \ref{disc}.

\subsection{Results at the Planck scale}

In \cite{modanese2019metrics} we chose units such that $c=\hbar=G=1$, and we chose to explore a duration and length scale $\tau=1$, $L=1$ (Planck scale). We took $N=100$ in order to have a meaningful but ``quick'' discretization and we run a Metropolis algorithm \cite{newman1999monte} with a return probability $\exp(-\beta^2 S^2)$ or $\exp(-\beta|S|)$ in order to avoid the instability problems related to the indefinite sign of the action.

The result of the simulations is that for suitable values of the inverse temperature $\beta$ one finds an ensemble of equilibrium configurations in which $\langle S \rangle \sim \delta \cdot 10^{-7} \sim 10^{-9}$ and $\langle S^2 \rangle \sim \delta^2 \cdot 10^{-14}$ or less.
The sum $\sum_{h=0}^N S_h$ is found to oscillate around zero with an amplitude $\sim 10^{-7}$.

In other words, after starting formally with a Lorentzian weight $e^{iS/\hbar}$ in order to skip the instability problems, and after realizing that the weight $e^{iS/\hbar}$ cannot be implemented numerically, the trick of using a weight $\exp(-\beta^2 S^2)$ or $\exp(-\beta|S|)$ in the algorithm is not meant as a solution of the instability or a way to study the full dynamics, but only as a way to obtain explicitly a set of fields with almost-zero action and a large volume in configuration space. 

In this context, the choice of the inverse temperature $\beta$ is a matter of convenience. After some trials we find that if $\beta$ is too small (high temperature) the algorithm stabilizes quickly but the configurations obtained have values of the action fluctuating in a wide range. On the other hand, if $\beta$ is too large (low temperature) equilibrium cannot be obtained in a reasonable number of steps and the system keeps ``drifting'' in some direction. In general, a Metropolis algorithm is in equilibrium at a certain temperature when the ratio between the frequency of steps towards lower energy/action and the frequency of steps towards higher energy/action remains approximately constant. As seen from Tab.\ \ref{table2}, however, we can fulfil the condition above by choosing $\beta$ in a certain range. For instance, for $N=12800$ we easily obtain thermalization in the range $128\cdot 10^7 \le \beta \le 1024\cdot 10^7$.

The $\{A_h\}$ configurations of the equilibrium ensemble have a peculiar dependence on the coordinate $r$ ($r=\delta h$): a sort of polarization with a step in the middle of the interval and $A<1$ in the inner region, $A>1$ in the outer region. Intuitively this matches the expectation of a cancellation between contributions to the integral of $R$ over different regions, which is also a typical feature of some of the analytical zero modes \cite{modanese2007vacuum}. One can check that the inner region has always negative $R$, while the opposite holds for the outer region. We shall see below that when the number $N$ of sub-intervals in the discretized action grows (continuum limit), this simple pattern of ``polarization into two regions'' changes.

\section{Scaling properties}
\label{sec-scaling}

Up to this point, what we can conclude is that at a length scale of the order of $L_P$ the discretized action permits the existence in the path integral of configurations which display strong deviations from flat space and polarization in $R$. In order to analyse the behavior at larger scale, we use natural units, in which $\hbar=c=1$. In these units the unit of length is 1 cm and the time $\tau$ is also expressed in cm$_t$, with the conversion 1 sec = 1 cm$_t \cdot (3\cdot 10^8)$. The Newton constant $G$ in natural units is equal to the square of the Planck length: $G \simeq L_P^2 \simeq 10^{-66}$ cm$^2$.

It follows that as magnitude order we can rewrite the discretized action (\ref{S-disc}) as
\begin{equation}
 \frac{S}{\hbar} \simeq 10^{66}\tau \hat{S}  \ \ \ \ ({\rm in \ natural \ units \ } \hbar=c=1)
\label{Ssuh}
\end{equation}
\begin{equation}
\hat{S}=\frac{L}{N} \sum_{h=0}^N S_h
\label{Shat}
\end{equation}

Due to the factor $10^{66}$, it seems very difficult to obtain $S/\hbar \ll 1$ in the discretized model, as soon as $L$ and $\tau$ are larger than the Planck scale $10^{-33}$ cm. If, however, we consider the continuum limit $N\to \infty$, it may be possible that the action $S/\hbar$ of the configurations built in our simulations becomes $\ll 1$, provided the scaling of $\hat{S}$ in $N$ is such that $\hat{S}\to 0$ quickly enough. For this reason we made several Metropolis simulations in order to compute $\hat{S}$ with increasing $N$.

\subsection{Results at ``macroscopic'' $L$}

In the first set of numerical trials that we performed in order to investigate the scaling for $N\to \infty$ we fixed the length $L$ to 10 units and the time $\tau$ to 1 unit. (Thinking of the field configurations in terms of vacuum fluctuations, $L$ represents their size and $\tau$ their duration.) The purpose of this choice is to compare and connect the present data, at the numerical level, with those obtained in our previous work. Here, however, the physical units employed are different and $L=10$ means $L=10$ cm. This is a macroscopic scale and certainly not our final target. We shall see that it is straightforward to pass from results at this scale to results at a length $L$ more appropriate to vacuum fluctuations (atomic and subatomic scale). That is because in the Metropolis algorithm employed the parameters $\beta$ and $L$ are multiplied by each other, so any reduction of $L$ can be compensated by an increase in the reciprocal temperature $\beta$ without affecting the thermal convergence of the algorithm.

In order to make contact with our previous data, let us start with a number of sub-intervals $N=100$ and proceed by repeatedly multiplying $N$ by 2. In this first series of trials we change $\beta$ in inverse proportion to $N$, in such a way that the factor $\beta L/N$ in the exponent of $\exp(-\beta |\hat{S}|)$ stays constant and optimal thermalization is achieved. The variation in $\hat{S}$ is due to the factor $1/N$ and to the fact that the sum $\sum_h S_h$ has more terms. The good news is that this finer subdivision leaves $\sum_h S_h$ almost unchanged and so $\hat{S}$ scales as $1/N$ (Tab.\ \ref{table1}).

\begin{figure}[h]
  \begin{center}
\includegraphics[width=7.0cm,height=5.1cm]{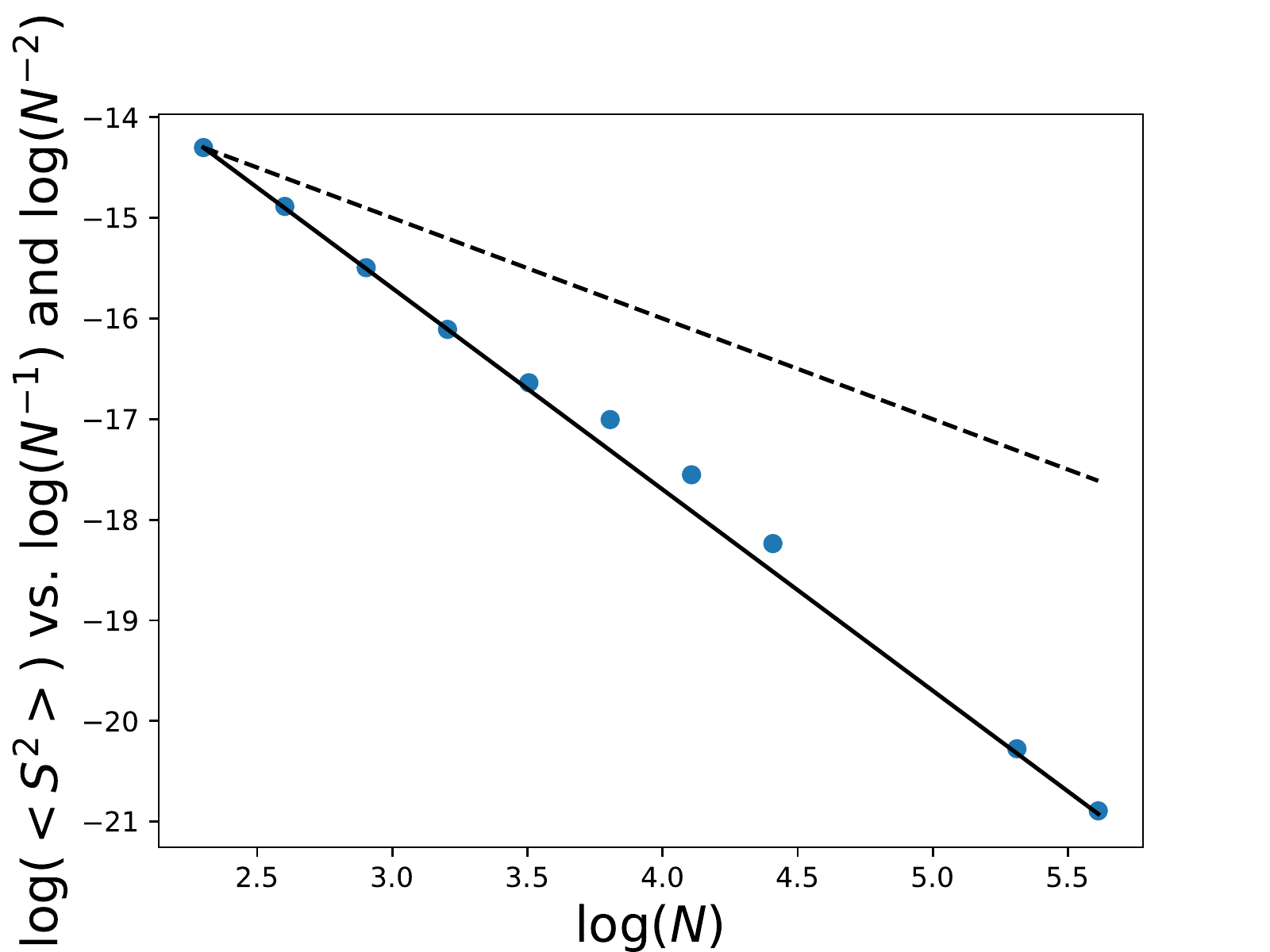}
\caption{Scaling of the average squared action $\langle S^2 \rangle$ as a function of $N$. Log-log scale; dots represent values from Tab.\ \ref{table1}, while the dashed line represents a dependence $N^{-1}$ and the solid line a dependence $N^{-2}$.
} 
\label{scaling}
  \end{center}
\end{figure}

When $N$ is increased, we also need to increase the number of Montecarlo steps in order to obtain precise results, because at each step the algorithm changes at random one value of $A_h$ in the range $h=0,1,...,N$, by an amount $\pm \varepsilon$. The averages in the table are computed well after thermalization (after 50\% of the steps for $N$ up to 25600, and then after 75\% of the steps). The quantity $\langle e^{-\beta \hat{S} } \rangle$ is the average of the return probability for the steps in which $|\hat{S}|$ increases. This probability increases with $N$ (except for $N=100$ and $N=200$, where the polarization pattern is changing, see Sect.\ \ref{pol}). $\langle \hat{S} \rangle$ and $\langle \hat{S}^2 \rangle$ are the averages of the action $\hat{S}$ and of its square. The most important quantity is $\langle \hat{S}^2 \rangle$, which tells us how much the action oscillates about its average. The dependence of $\langle \hat{S}^2 \rangle$ on $N$ reported in Tab.\ \ref{table1} is also plotted in Fig.\ \ref{scaling}, from which a scaling very close to $1/N^2$ can be deduced. This means that in the continuum limit $N\to \infty$ the phase $S/\hbar$, eq.\ (\ref{Ssuh}), can actually become $\ll 1$, in spite of the large dimensional factor $10^{66}$, especially if we start from microscopic values of $L$ and $\tau$ (see below). Moreover, there is a favourable scaling in the parameter $\beta$ (Tab.\ \ref{table2}).

How should one interpret the increasing values of $\beta$ needed for equilibrium as $N$ grows? As explained at the beginning of Sect.\ \ref{sec-dis}, lower values of $\beta$ imply in general larger fluctuations of the action, while we need $\langle \hat{S}^2 \rangle$ to decrease at least as $1/N^2$ for the continuum limit to be effective. The interpretation is then that in the continuum limit the field configurations of the equilibrium ensemble have a very low temperature, i.e.\ they are very close to the minimum of the classical action. This is in qualitative agreement with the findings of \cite{bonanno2013modulated}, namely that the true minimum of the stabilized action is obtained for a class of oscillating metrics, breaking translational invariance.

\begin{table}
\begin{center}
\begin{tabular}{|c|c|c|c|c|c|} 
\toprule
$N$ &\ $\ \beta \ $ \ &\ MC steps \ &  $\langle e^{-\beta |\hat{S}|} \rangle$ \ & $\langle \hat{S} \rangle$ & $\langle \hat{S}^2 \rangle$ \nonumber \\
\hline
100 & $10^7$ &  $2\cdot 10^9$ & 0.17 & $4.3\cdot 10^{-9}$ & $2.0 \cdot 10^{-14}$ \nonumber \\
200 & $2\cdot 10^7$ &  $2\cdot 10^9$ & 0.11 & $6.4\cdot 10^{-9}$ & $5.1 \cdot 10^{-15}$ \nonumber \\
400 & $4\cdot 10^7$ & $4\cdot 10^9$ & 0.022 & $5.4\cdot 10^{-9}$ & $1.3 \cdot 10^{-15}$  \nonumber \\
800 & $8\cdot 10^7$ & $8\cdot 10^9$ & 0.024 & $1.4\cdot 10^{-9}$ & $3.2 \cdot 10^{-16}$  \nonumber \\
1600 & $16\cdot 10^7$ & $8\cdot 10^9$ & 0.055 & $3.0\cdot 10^{-10}$ & $7.8 \cdot 10^{-17}$ \nonumber \\
3200 & $32\cdot 10^7$ & $8\cdot 10^9$ & 0.14 & $1.2\cdot 10^{-9}$ & $2.3 \cdot 10^{-17}$ \nonumber \\
6400 & $64\cdot 10^7$ & $8\cdot 10^9$ & 0.27 & $1.5\cdot 10^{-9}$ & $9.9 \cdot 10^{-18}$  \nonumber \\
12800 & $128\cdot 10^7$ & $16\cdot 10^9$ & 0.25 & $8.7\cdot 10^{-10}$ & $2.8 \cdot 10^{-18}$ \nonumber \\
25600 & $256\cdot 10^7$ & $16\cdot 10^9$ & 0.29 & $3.4\cdot 10^{-10}$ & $5.8 \cdot 10^{-19}$ \nonumber \\
204800 & $2048\cdot 10^7$ & $16\cdot 10^{10}$ & & $3.3\cdot 10^{-11}$ & $5.3 \cdot 10^{-21}$ \nonumber \\
409600 & $4096\cdot 10^7$ & $16\cdot 10^{10}$ & & $1.5\cdot 10^{-11}$ & $1.3 \cdot 10^{-21}$ \nonumber \\
\hline
\end{tabular}	
\caption{Average values of the return probability
$\langle e^{-\beta \hat{S}} \rangle$, the action $\langle \hat{S} \rangle$ and the squared action $\langle \hat{S}^2 \rangle$ in dependence on $N$
(number of sub-intervals of $(0,L)$). Here $L=10$ cm. The inverse temperature $\beta$ changes in proportion to $N$, in order to maintain the factor $\beta/N$ constant. The discretized field components $A_h$ are randomly increased in the Montecarlo steps by $\pm \varepsilon$, with $\varepsilon=10^{-6}$. 
For the last two values of $N$ the calculation of $\langle e^{-\beta \hat{S}} \rangle$ (and of $\langle A_h \rangle$) was omitted in order to speed-up the algorithm and increase the precision in $\langle \hat{S}^2 \rangle$.
See also plot of $\langle \hat{S}^2 \rangle$ in Fig.\ \ref{scaling}.
}		
\label{table1}
\end{center}
\end{table}

\begin{table}
\begin{center}
\begin{tabular}{|c|c|c|c|c|c|} 
\toprule
$\ \beta \ $ &\ $N$ \ &\ MC steps \ &  $\langle e^{-\beta |\hat{S}|} \rangle$ \ & $\langle \hat{S} \rangle$ & $\langle \hat{S}^2 \rangle$ \nonumber \\
\hline
$128\cdot 10^7$ & 12800 & $16\cdot 10^9$ & 0.25 & $8.7\cdot 10^{-10}$ & $2.8 \cdot 10^{-18}$ \nonumber \\
$256\cdot 10^7$ & 12800 & $16\cdot 10^9$ & 0.16 & $5.0\cdot 10^{-10}$ & $8.2 \cdot 10^{-19}$ \nonumber \\
$512\cdot 10^7$ & 12800 & $16\cdot 10^9$ & 0.095 & $3.3\cdot 10^{-10}$ & $2.8 \cdot 10^{-19}$ \nonumber \\
$1024\cdot 10^7$ & 12800 & $16\cdot 10^9$ & 0.055 & $2.2\cdot 10^{-10}$ & $1.0 \cdot 10^{-19}$ \nonumber \\
\hline
\end{tabular}	
\caption{Scaling of $\langle e^{-\beta \hat{S}} \rangle$, $\langle \hat{S} \rangle$ and $\langle \hat{S}^2 \rangle$ in dependence on $\beta$ with $N$ fixed, $L=10$, $\varepsilon =10^{-6}$. Note the decrease of the average return probability $\langle e^{-\beta \hat{S}} \rangle$, coherent with the role of the inverse temperature $\beta$ in the thermalization process.}		
\label{table2}
\end{center}
\end{table}

\subsection{Scaling in $\beta$ for microscopic $L$}

Tab.\ \ref{table3} shows the scaling of $\langle e^{-\beta \hat{S}} \rangle$, $\langle \hat{S} \rangle$ and $\langle \hat{S}^2 \rangle$ in dependence on $\beta$ for $L$ at a microscopic scale, namely $L =10^{-13}$ cm.

The first value of $\beta$ is chosen, to ensure thermalization, in such a way that the product $L\beta$ is the same as for data with $L=10$, $\beta=128\cdot 10^7$, $N=12800$; this implies that $\beta$ must now be equal to $128\cdot 10^{21}$.

The results for $\langle \hat{S} \rangle$ and $\langle \hat{S}^2 \rangle$ are seen in Tab.\ \ref{table3} to scale in proportion to $L$, in comparison to the results in Tab.\ \ref{table1}. This could have been predicted from the fact that the discretized action is proportional to $\delta=L/N$, and so are its variations $\Delta \hat{S}$ and $\Delta |\hat{S}|$ in the Montecarlo algorithm. We obtain here another confirmation that the algorithm scales as expected with respect to the parameters $L$ and $\beta$.

\begin{table}
\begin{center}
\begin{tabular}{|c|c|c|c|c|c|} 
\toprule
$\ \beta \ $ &\ $N$ \ &\ MC steps \ &  $\langle e^{-\beta |\hat{S}|} \rangle$ \ & $\langle \hat{S} \rangle$ & $\langle \hat{S}^2 \rangle$ \nonumber \\
\hline
$128\cdot 10^{21}$ & 12800 & $16\cdot 10^9$ & 0.25 & $8.8\cdot 10^{-24}$ & $2.9 \cdot 10^{-46}$ \nonumber \\
$256\cdot 10^{21}$ & 12800 & $16\cdot 10^9$ & 0.15 & $4.9\cdot 10^{-24}$ & $8.0 \cdot 10^{-47}$ \nonumber \\
$512\cdot 10^{21}$ & 12800 & $16\cdot 10^9$ & 0.093 & $3.3\cdot 10^{-24}$ & $2.8 \cdot 10^{-47}$ \nonumber \\
$1024\cdot 10^{21}$ & 12800 & $16\cdot 10^9$ & 0.053 & $2.2\cdot 10^{-24}$ & $1.0 \cdot 10^{-47}$ \nonumber \\
\hline
\end{tabular}	
\caption{Scaling of $\langle e^{-\beta \hat{S}} \rangle$, $\langle \hat{S} \rangle$ and $\langle \hat{S}^2 \rangle$ in dependence on $\beta$ with $N$ fixed, $L =10^{-13}$ cm (``microscopic scale''), $\varepsilon =10^{-6}$.}		
\label{table3}
\end{center}
\end{table}

\subsection{Polarization pattern}
\label{pol}

\begin{figure}[h]
    \includegraphics[width=10.5cm,height=6.1cm]{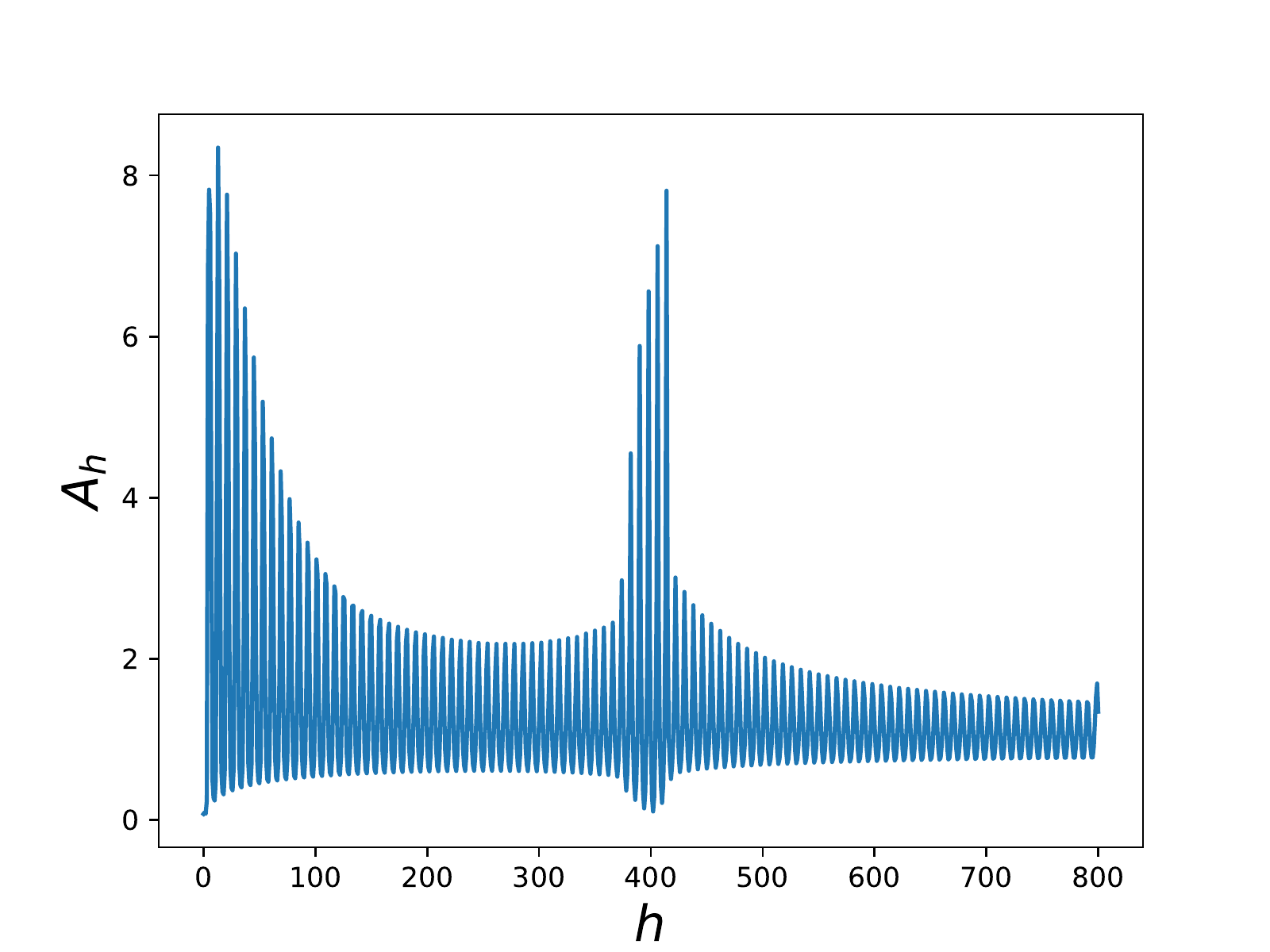}
    \includegraphics[width=10.5cm,height=6.1cm]{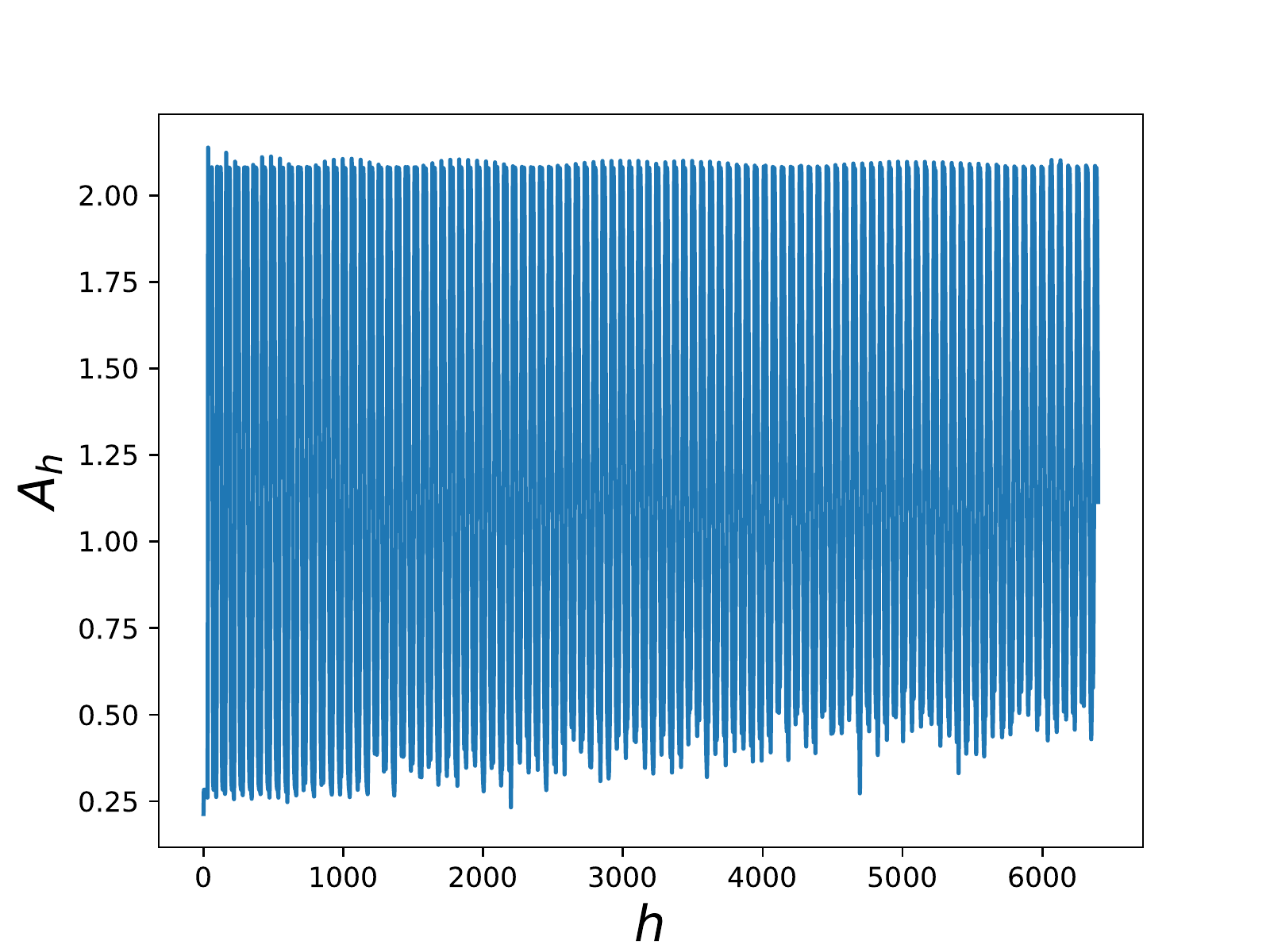}
\caption{(a) Polarization pattern for $N=800$ sub-intervals. 
(b) Polarization pattern for $N=6400$ sub-intervals. 
} 
\label{pol-1}
\end{figure}

\begin{figure}[h]
    \includegraphics[width=10.5cm,height=6.1cm]{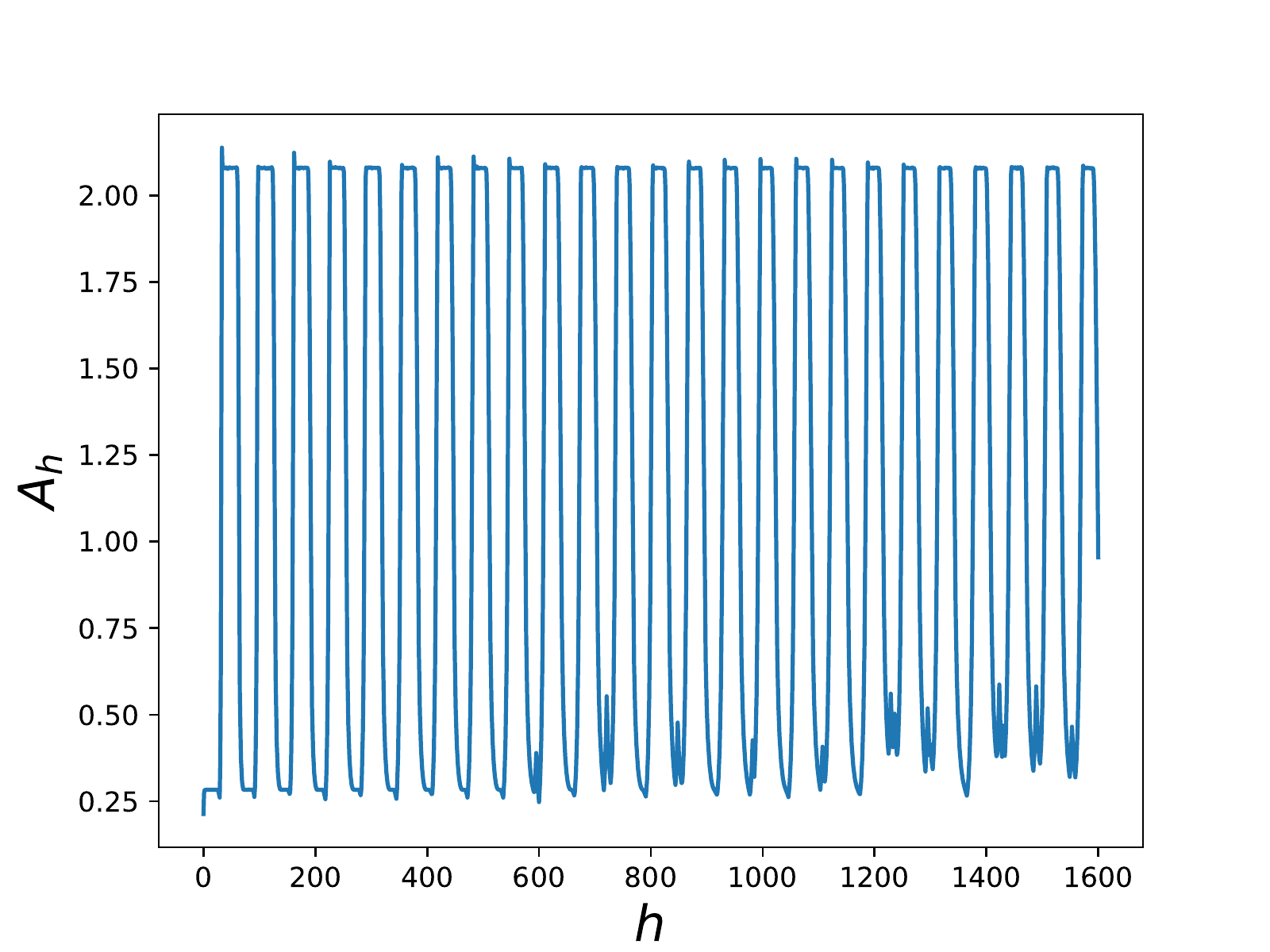}
    \includegraphics[width=10.5cm,height=6.1cm]{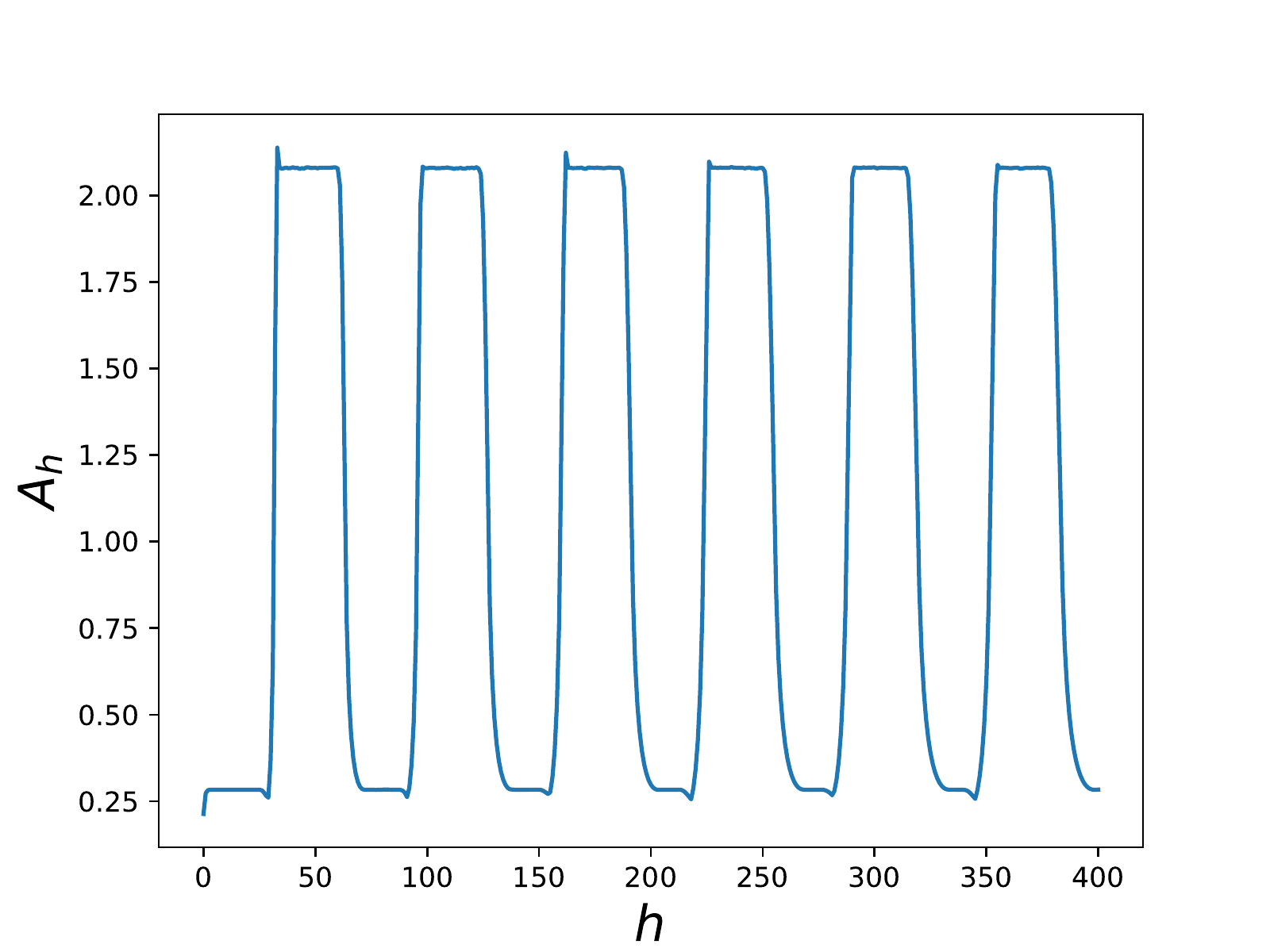}
\caption{(a) detail of the polarization pattern for $N=6400$ sub-intervals, showing $1/4$ of the total interval. 
(b) Showing $1/16$ of the total interval. 
} 
\label{pol-2}
\end{figure}

The simple bipolar polarization pattern observed in the averaged field values $\langle A_h \rangle$ for $N=100$ \cite{modanese2019metrics} changes when $N$ increases. Multiple oscillations begin to appear, with an envelope changing with $N$ (see an example in Fig.\ \ref{pol-1}, (a)), until for $N$ approximately greater than 3200 the situation stabilizes and all the oscillations have almost exactly the same amplitude (Fig.\ \ref{pol-1}, (b)). The number of oscillations does not depend on any of the physical parameters $N$, $L$, $\beta$ and $\varepsilon$. It appears to be a general ``mathematical'' feature of the minimum configuration of the discretized action $\sum_h S_h$. Fig.\ \ref{pol-2} shows two details of Fig.\ \ref{pol-1} (b), namely on sub-intervals with 1600 and 400 values of $h$. From these details we can see that the fixed total number of oscillations in the interval $(0,L)$ is approximately equal to $10^2$, even though, as mentioned, there appears to be no relation between this number and the physical parameters.

As discussed in \cite{modanese2019metrics}, in polarized configurations of this kind there are contributions to the local curvature coming both from the plateaus and from the steps. When $N$ is large, the plateaus comprise hundreds of values of $h$ (the discretization index). For each $h$ we have a value of $\langle A_h \rangle$ at the end of the simulation, and on the plateaus these values are typically constant up to $10^{-3}$. (The values of $A_h$ displayed in Figs.\ \ref{pol-1}, \ref{pol-2} are actually averages $\langle A_h \rangle$.) The steps comprise only a few values of $h$, with standard deviation of $\langle A_h \rangle$ of the order of $10^{-2}$. This shows that after the Monte Carlo algorithm has attained thermal equilibrium, this equilibrium is quite stable. As displayed in Tab.\ \ref{table1}, when $N$ increases thermalization requires a lower temperature.

\begin{figure}[h]
    \includegraphics[width=7.0cm,height=5.1cm]{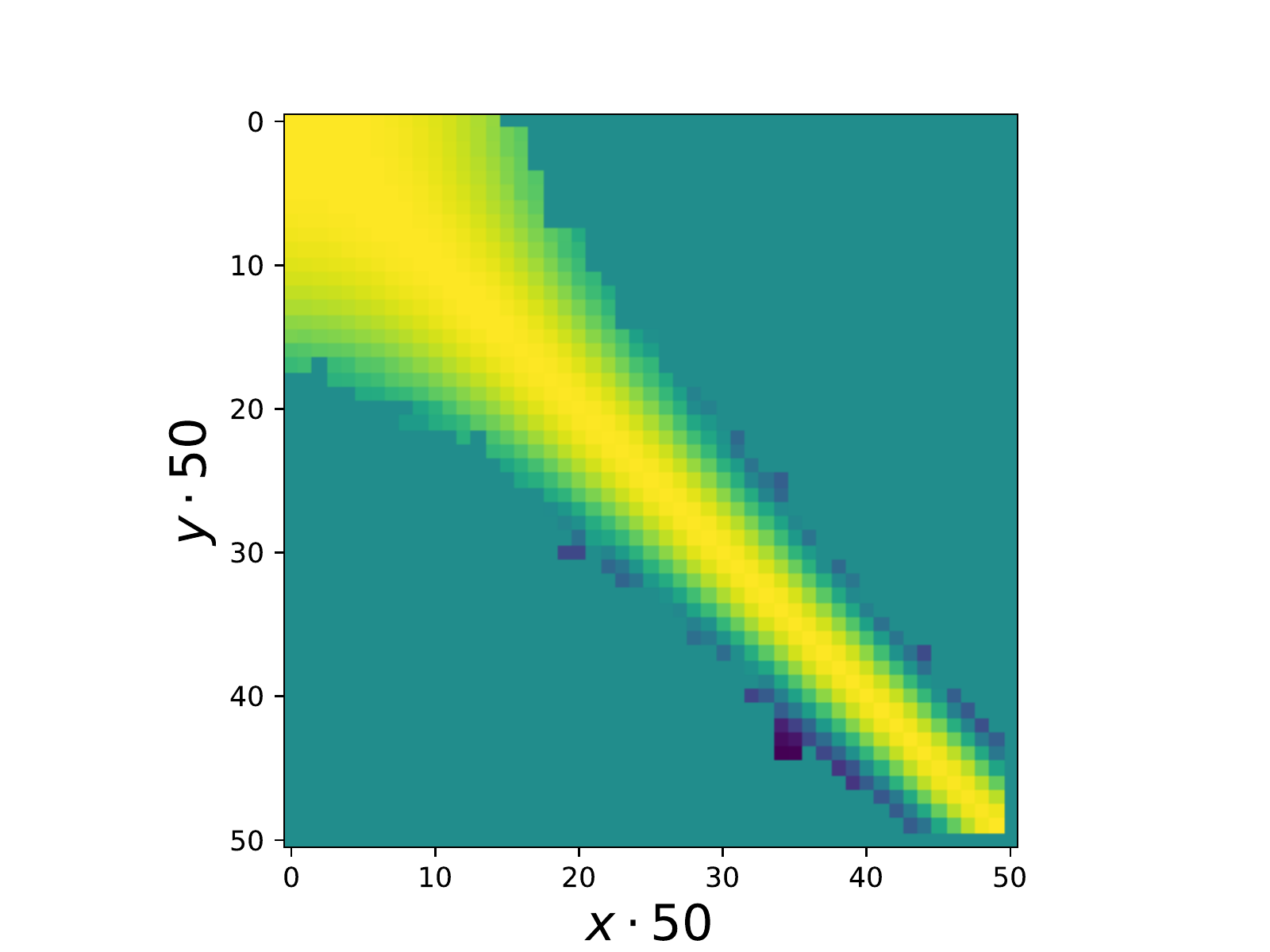}
  \includegraphics[width=7.0cm,height=5.1cm]{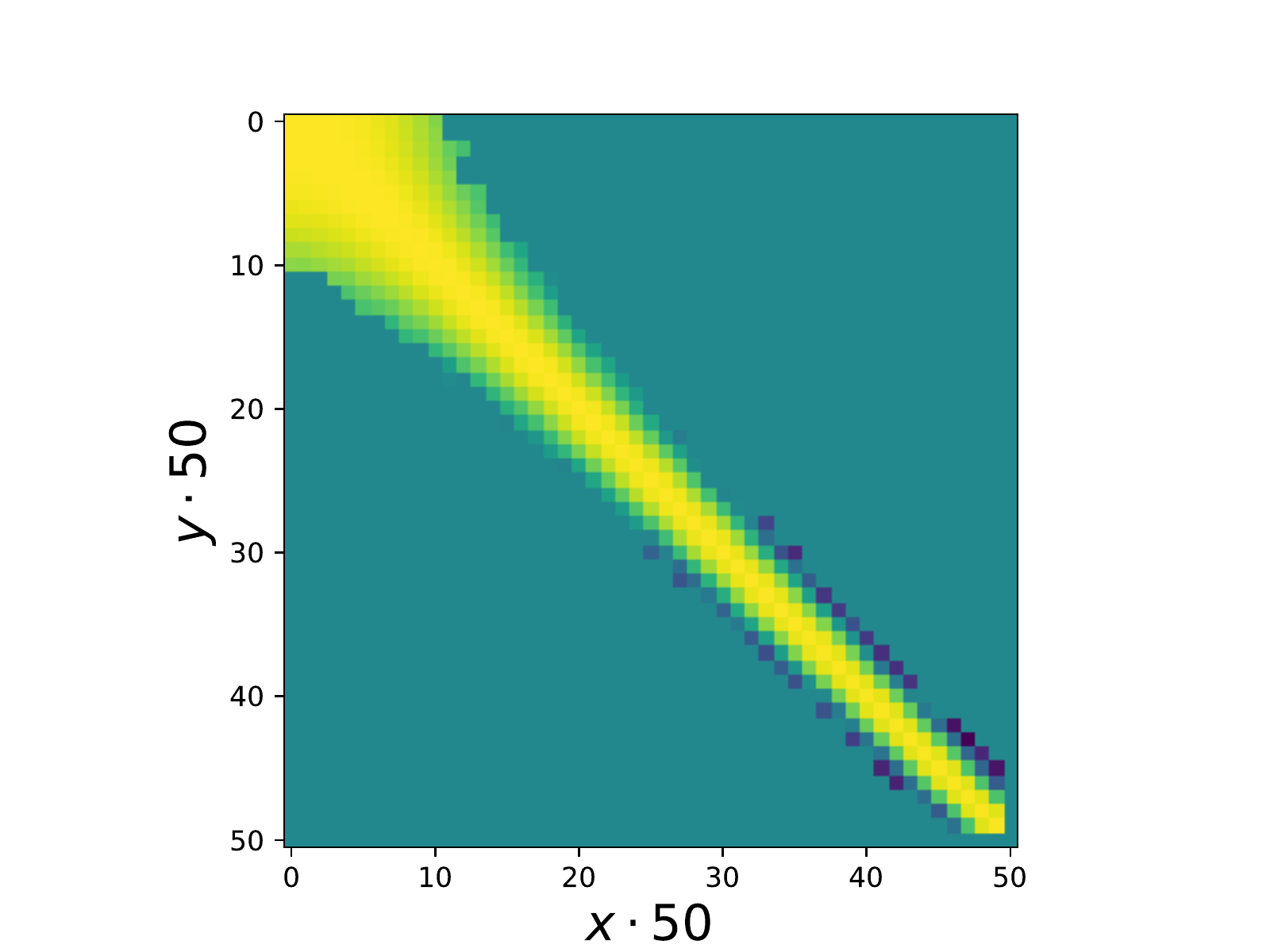}
    \includegraphics[width=7.0cm,height=5.1cm]{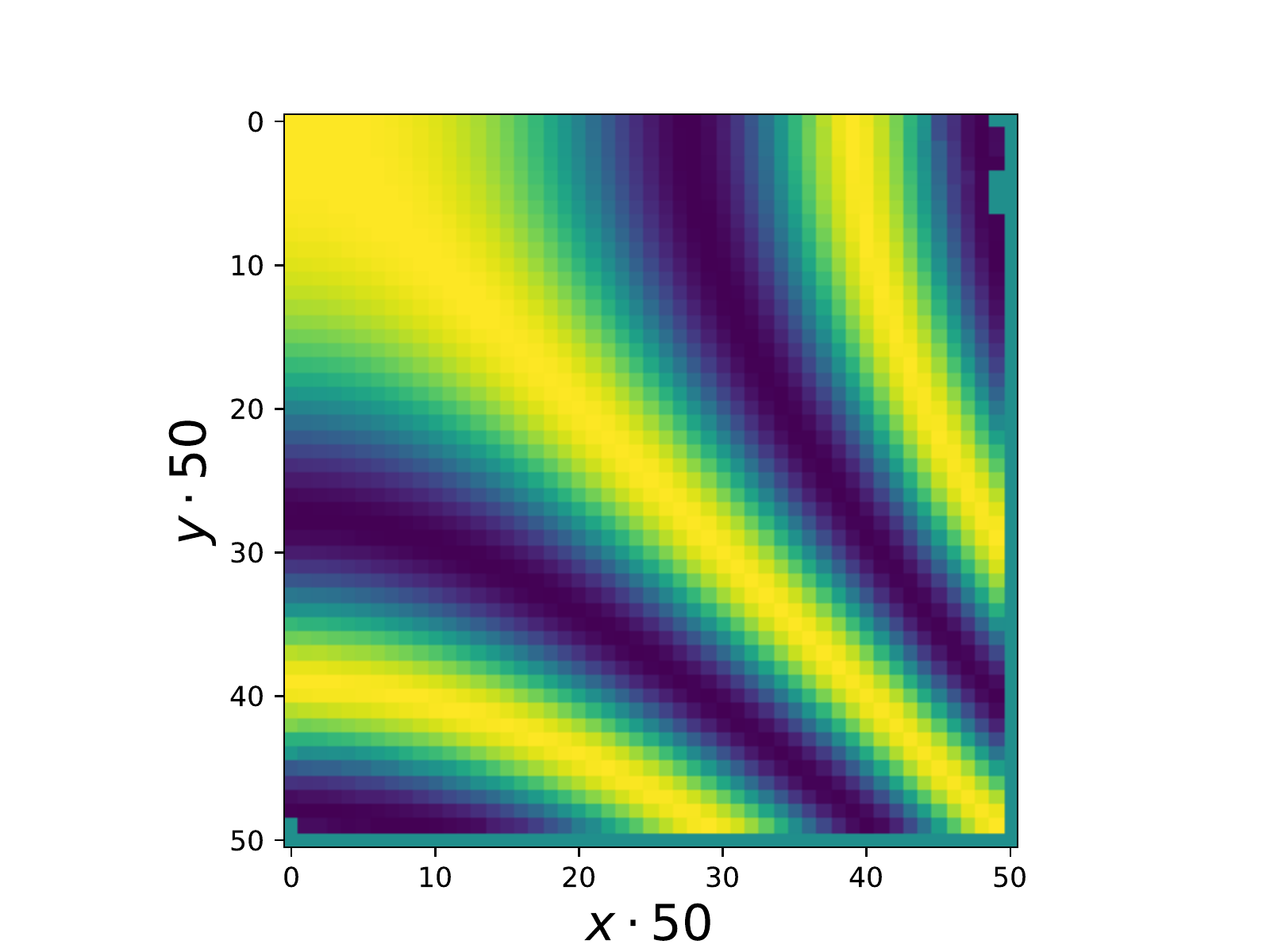}
\caption{Contributions to the integral of $\cos[\hbar^{-1}(x^2-y^2)]$ in the square $\{0\le x\le 1, 0\le y \le 1\}$ obtained through an adaptive Monte Carlo integration with inverse temperature $\beta$. Parameters: (a) $\hbar=0.1$, $\beta=1$. (b) $\hbar=0.05$, $\beta=2$. (c) $\hbar=0.1$, $\beta=7.8\cdot 10^{-3}$. The contributions near the origin come from the stationary point of the phase, those along the diagonal from the zero mode $y=x$. In (b) the region near the stationary point is smaller because $\hbar$ is smaller, but the length of the zero mode is unaffected. In (c) the contributions of the disconnected zero modes $y=x\pm 2\pi \hbar$ also appear, because the temperature is much higher. 
} 
\label{staz-1}
\end{figure}

\section{Discussion, conclusions}
\label{disc}

\subsection{A 2D integral with stationary phase and zero modes}
\label{sec-2di}

A key concept of this work, already discussed analytically in \cite{modanese2007vacuum,modanese2016functional,modanese2017ultra}, is that of zero modes of the action. This relates to a peculiar property of the gravitational field, not easily found in other physical systems: the non-positivity of the action in the path integral $\int \, d[g_{\mu \nu}]e^{iS/\hbar}$. 

A simple mathematical example can help to elucidate the idea of zero modes. Consider a 2D integral with oscillating integrand, of the form
\begin{equation}
I=\int dx \int dy \cos[\phi(x,y)]f(x,y),\ \ \ {\rm with} \ \ \phi(x,y)=\hbar^{-1}(x^2-y^2)
\label{integrale}
\end{equation}
where $f(x,y)$ is a smooth function, and suppose that $\hbar \ll 1$. We expect the main contribution to the integral to come from the region near the origin $x=0$, $y=0$, where the phase of the cosine is stationary. This would in fact be true if the phase was $\phi=[\hbar^{-1}(x^2+y^2)]$. However in this case the phase is zero, even if not stationary, along the lines $y=\pm x$. Could the infinite region along these lines give a contribution to the integral comparable to the region near the origin? This can be verified using an algorithm for numerical integration similar to an adaptive Monte Carlo. The algorithm samples the integrand at random starting from the origin and moving in small steps $(\delta x$, $\delta y)$. Each step is accepted unconditionally if it gives an increase $\delta |\cos \phi|$ positive, or else accepted with probability $e^{\beta \delta |\cos \phi|}$. In this way, the sampling points are more dense in the regions where there are larger contributions to the integral, and the effect can be tuned varying the inverse temperature $\beta$.

Let us reduce the integration region to the square $\{ 0\le x \le 1,\ 0\le y \le 1\}$ and divide it into, for example, $50 \times 50$ cells of side $a=0.02$ with indices $i$, $j$. If the number of sampling points falling in the cell $(i,j)$ is $c_{ij}$ and the sum of the values of the integrand at those points is $s_{ij}$, the integral is approximated by 
\[
I \simeq a^2 \sum_{i,j} \frac{s_{ij}}{c_{ij}}
\]

Fig.\ \ref{staz-1} represents with a density plot the contributions of the individual cells in a case where the function is simply $f(x,y)=1$ (see caption for details). If we compute instead the average $\langle r \cos \phi \rangle$ over all sampling points, namely with $f(x,y)= \sqrt{x^2+y^2} $, we obtain at low temperature approximately 0.7 (half the diagonal), showing that the regions which contribute to the integral are in fact spread along the zero mode. However, when the temperature is increased (Fig.\ \ref{staz-1}, (c)) the strong destructive interference along the zero modes tend to cancel their contributions, leaving only the contribution near the origin. This can also be seen from the fact that the average $\langle r \cos \phi \rangle$ decreases.

In the simple case of the 2D integral in $x$, $y$ of eq.\ (\ref{integrale}) 
 all these properties can be easily predicted, because we can plot the integrand and we know that the main contributions arise in the regions where the integrand is large and are directly proportional just to the area of these regions. One can also predict that being zero modes 1-dimensional, in the limit of small $\hbar$ they do not contribute to the 2D integral. 

\subsection{Extension to higher dimension}
\label{sec-ext}

For a path integral in infinite dimensions, with a non-polynomial action, all this {\it a priori} information is not available. Even if we are able to solve the exact equation for the zero modes (analogue of $x^2-y^2=0$ in the 2D example) \cite{modanese2007vacuum,modanese2019metrics}, it is hard to assess the ``volume'' of the solutions in the functional space, and even harder to asses this volume for the weaker but crucial condition $S\ll \hbar$.

An higher-dimensional extension of the polynomial example above could be in principle the following: consider the integral
\[
\int d^mx \int d^ny \, \cos[\hbar^{-1}(x_1^2+...+x_m^2-y_1^2-...-y_n^2)]
\]
and the zero modes of the phase, which satisfy the equation
\[
x_1^2+...+x_m^2-y_1^2-...-y_n^2=0
\]
The dimension of these modes is $(m+n-1)$ (for example, for a phase proportional to $(x_1^2+x_2^2-y_1^2)$ the zero mode is a conical surface), so for $m,n\to \infty$ they might indeed contribute to the integral.

To complete the analogy, note that in the gravitational case the contribution to the adaptive Monte Carlo coming from the configurations which make the action stationary appears to be actually negligible.

\subsection{Limitations of the present approach and comparison with other methods}
\label{sec-limits}

The use of the absolute value of the action in the Euclidean path integral allows to circumvent the stability issues. It is admittedly a strong assumption, whose validity should be further checked, and which does not hold for the dynamics of configurations with action $|S|\gg \hbar$, such that the phase factor in the path integral is rapidly oscillating.

For the practical purposes of a numerical simulation in the region $|S|\ll \hbar$, however, using the absolute value appears to be not very different from other stabilization techniques, like the introduction of an $R^2$ term \cite{bonanno2013modulated}. Being the flat space configuration, with $R=0$ everywhere, a stationary point, the absolute value does not produce any discontinuity in the derivatives of the action. In previous versions of the simulations we used the squared action instead of the absolute value, obtaining similar results. The algorithm could be further adapted to the insertion of an $R^2$ term.

One can safely state, in any case, that results of simulations with the absolute value are exact for a theory with action $|S_{E.H.}|$, which does not coincide in general with the Einstein-Hilbert theory, but has the same classical field equations, obtained minimizing $|S_{E.H.}|$. The logic here would be similar to that of theories with lagrangian $f(R)$ \cite{nojiri2017modified}, even though at the level of perturbative quantum field theory only the Einstein-Hilbert action represents massless particles with spin 2.

On another front, we note that in the present approach it is impossible to address the diffeomorphism symmetry as clearly as done, for instance, in the Regge calculus with full simulation of the quantum dynamics (\cite{hamber2008quantum,hamber2019vacuum} and refs.). Even in the spherically symmetric case presented here, an invariance under reparametrisations of the coordinate $r$ remains. When one generates a new metric configuration, in principle it is possible that the geometry is not actually being changed, but one is just doing such a reparametrization. In practice, however, this is extremely unlikely when randomly changing one of the discrete variables $A_h$ at a time, as it happens in our algorithm. Furthermore, {\it a posteriori} we can be sure that the configurations found at equilibrium are really different from the flat space we started from.

In these configurations, the coordinate distance cannot be interpreted as a physical distance, the latter being given instead by the usual expression $ds^2=g_{rr}dr^2$. This means, for instance, that the real length of the upper plateaus in Fig.\ \ref{pol-2} is definitely larger than the length of the lower plateaus.

The analogies between our results and those of Ref.\ \cite{bonanno2013modulated} are stimulating, but several differences should be noticed, in addition to the different stabilization methods: 

(1) In \cite{bonanno2013modulated}, the degree of freedom in the metric is a conformal factor, while here it is the component $g_{rr}$ in a stationary approximation. 

(2) The authors of \cite{bonanno2013modulated} search for the vacuum state by minimizing the action through a rigorous analytical approach, while we rely on numerical simulations. That is why they interpret the rippled spacetime obtained as one that becomes flat
upon averaging over a periodicity volume, i.e.\ after a purely {\it classical} coarse graining. In our discretized model, we interpret the rippled spacetime obtained as an ensemble of purely quantum states with no classical counterpart. We also find, however, that a proper continuum limit is possible only in the low temperature limit (large $\beta$), which takes us back to the classical theory. In this sense there is a qualitative agreement between the two approaches. There might also be a connection between our concept of zero modes of the action and the restricted-space minimization of Sect.\ 2 in \cite{bonanno2013modulated}.

\subsection{Conclusion}
\label{sec-concl}

In this article, we have studied the gravitational path integral of spherically symmetric space-times independent in time using numerical Monte Carlo methods. The system is first reduced to the spherically symmetric and time-independent setting, before it is discretized in radial direction. The reduced Einstein-Hilbert action is Wick-rotated and only its absolute value is considered in the path integral, since it is not bounded from below. The goal is to explore the configurations with almost vanishing action since these might contribute in the full Lorentzian path integral. In the numerical studies oscillations are found that suggest large deviations from the classical vacuum solution.

Usually, one expects path integrals to be dominated by classical solutions given appropriate boundary data. However, this does not seem to be the case here as typical configurations appear to significantly deviate from the classical vacuum solution and show oscillatory behaviour. The interpretation of these configurations is not completely clear and open questions remain on whether it would be realistic to find them in the full Lorentzian path integral and, if yes, if this would result in observational consequences.

A non-perturbative Monte Carlo algorithm for the discretized action like that employed in this work (and in much more complete form by Hamber, Ambj{\o}rn and co-workers \cite{hamber2008quantum,hamber2019vacuum,ambjorn2012nonperturbative,loll2019quantum}) seems to be at present one of the best tools available for exploring  quantum metrics closely connected to the classical vacuum state like the polarized configurations we have found in this work. The astounding detailed structure of these configurations (Sect.\ \ref{pol}) and their stability and reproducibility are intriguing, and possibly part of more general patterns valid beyond the approximations made here (spherical symmetry, $g_{00}=1$, modes almost stationary in time). 

Some physical comparisons can be drawn, as in \cite{bonanno2013modulated,bonanno2019structure}, to kinetic condensates in other quantum field theories \cite{lauscher2000rotation} and to anti-ferromagnetic systems in statistical physics (\cite{branchina1999antiferromagnetic} and refs.).

We have shown that the average squared action $\langle \hat{S}^2 \rangle$ of the polarized configurations scales as $1/N^2$ up to a number $N$ of sub-intervals of the order of $10^6$, for any length scale $L$. If this behavior can be extrapolated to larger $N$, their adimensional action $S/\hbar\simeq 10^{66}\tau \hat{S}$ can be $\ll 1$ also at scales much larger than the Planck scale. 

Future work should be devoted to an extension of the simulations to the case with angular and time dependence, and to a phenomenological comparison with observational constraints on gravitational vacuum fluctuations \cite{amelino2001phenomenological,quach2015gravitational}.

\bibliographystyle{unsrt}
\bibliography{QG2}

\end{document}